\PassOptionsToPackage{square,numbers}{natbib}
\documentclass[twocolumn,twoside,slac_two]{revtex4}
\usepackage[square,numbers]{natbib}
\usepackage{graphicx}
\usepackage{fancyhdr}
\newcommand{\mnras}{MNRAS}
\newcommand{\apjl}{ApJL}
\newcommand{\apj}{ApJ}
\newcommand{\aap}{A\&A}

\newcommand{\reffig}[1]{Fig.~\ref{#1}}
\newcommand{\refpar}[1]{\S\ref{#1}}

\newcommand{\refeq}[1]{Eq.(\ref{#1})}

\begin{document}

%%%%%%%%%%%%%%%%%%%%%%%%%%%%%%%%%%%%%%%%%%%
\title{Consequences of the $\gamma$$\gamma$ attenuation in GRBs: a detailed study}
\author{R. Hasco\"et, F. Daigne$^*$, R. Mochkovitch, V. Vennin}
\affiliation{UPMC - CNRS, UMR 7095, Institut d'Astrophysique de Paris, F-75014, Paris, France\\
$^*$Institut Universitaire de France}

%%%%%%%%%%%%%%%%%%%%%%%%%%%%%%%%%%%%%%%%%%%
\begin{abstract}
Recent detections of GeV photons in a few GRBs by Fermi-LAT have led to strong constraints on the bulk Lorentz factor in GRB outflows. To avoid a large $\gamma$$\gamma$ optical depth, minimum values of the Lorentz factor have been estimated to be as high as 800-1200 in some bursts. Here we present a detailed calculation of the $\gamma$$\gamma$ optical depth taking into account both the geometry and the dynamics of the jet. In the framework of the internal shock model, we compute lightcurves in different energy bands and the corresponding spectrum and we show how the limits on the Lorentz factor can be significantly lowered compared to previous estimates.
 
Our detailed model of the propagation of high energy photons in GRB outflows is also appropriate to study many other consequences of $\gamma$$\gamma$ annihilation in GRBs: (i) the $\gamma$$\gamma$ cutoff transition in a time-integrated spectrum is expected to be closer to a power-law steepening of the spectrum than to a sharp exponential decay; (ii) the temporal evolution of the $\gamma$$\gamma$ opacity during a burst favors a delay between the MeV and GeV light curves; (iii) for complex GRBs, the $\gamma$$\gamma$ opacity suppresses the shortest time-scale features in high energy light curves (above 100 MeV). Finally we also consider GRB scenarii where MeV and GeV photons are not produced at the same location, showing that the $\gamma$$\gamma$ opacity could be further lowered, reducing even more the constraint on the minimum Lorentz factor.
\end{abstract}
%%%%%%%%%%%%%%%%%%%%%%%%%%%%%%%%%%%%%%%%%%%

\maketitle
\thispagestyle{fancy}

%%%%%%%%%%%%%%%%%%%%%%%%%%%%%%%%%%%%%%%%%%%
\section{Introduction}
\textbf{The compactness problem.} 
The short time scales observed in GRBs (down to a few ms) can be used to deduce an upper limit on the size of the emitting region producing $\gamma$-rays. This information combined with the huge isotropic $\gamma$-ray luminosities deduced from the measured redshifts imply huge photon densities. Then the simplest assumption of an emission produced by a plasma radiating isotropically with no macroscopic motion predicts that $\gamma$-ray photons should not escape due to $\gamma$$\gamma$ annihilation $\gamma \gamma \rightarrow e^+ e^-$. This is in contradiction with the observed GRB spectra which are non-thermal and extend well above the rest-mass electron energy $m_e c^2 \approx 511$ keV.
Observation and theory can be reconciled by assuming that the emitting material is moving at ultra-relativistic velocities \cite{rees:1966}. This is mainly due to the relativistic beaming.  First it implies that the observer will see only a small fraction of the emitting region: the constraint on the size of the source is now less severe. Second, the collimation of photons in the same direction reduce the number of potential interactions. Finally the typical $\gamma$$\gamma$ interaction angle becoming small the photon energy threshold for pair production becomes higher. This theoretical context combined with the observational data gives the possibility to estimate a minimum Lorentz factor $\Gamma_{\mathrm{min}}$ for the emitting outflow in 
GRBs \cite{lithwick:2001} or directly a Lorentz factor estimate if the $\gamma$$\gamma$ cutoff is clearly identified in the spectrum (see \cite{ackermann:2011}).
\newline 

\noindent \textbf{Severe constraints on the Lorentz factor from Fermi-LAT observations.} 
Since the launch of Fermi in June 2008, the LAT instrument has detected high energy photons above 10 GeV in a few GRBs. The observed $\gamma$-ray spectrum often remains consistent with a Band function covering the GBM and LAT spectral ranges without any evidence of a high energy cutoff that could be identified as a signature of $\gamma \gamma \rightarrow e^+ e^-$. This extension by Fermi of the observed spectral range upper bound from a few MeV (e.g. BATSE) to 10 GeV implies constraints on $\Gamma_{\mathrm{min}}$ which are much more severe than the ones obtained previously. In a few cases $\Gamma_{\mathrm{min}}$ has been estimated to be of the order of 1000 (for example: GRB 080916C -- $\Gamma_{\mathrm{min}}$ = 887   \cite{abdo:2009a}, GRB 090510 -- $\Gamma_{\mathrm{min}}$ = 1200 \cite{ackermann:2010}). These extreme values put severe constraints on the physics of the central engine which should be able to strongly limit the baryon load in the outflow. 

\noindent However these $\Gamma_{\mathrm{min}}$ values were obtained from a simplified ``single zone'' model where the space and time dependencies are averaged out. The motivation of this work is to develop a detailed approach taking  into account a more realistic treatment of the dynamics.

\section{Computing the $\gamma$$\gamma$ optical depth}
\label{section_computing}

\begin{figure}
\centerline{\includegraphics[width=0.45\textwidth]{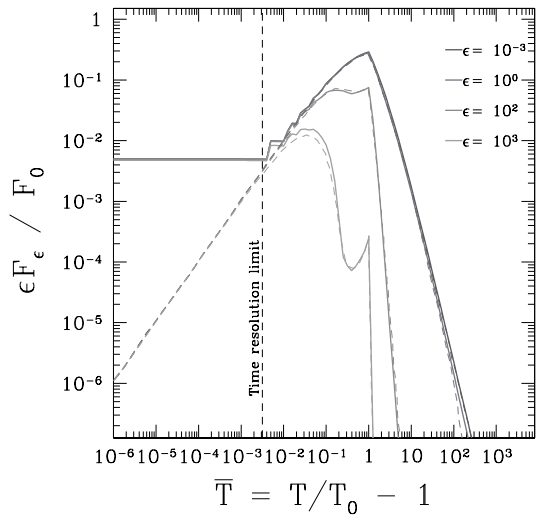}}
\caption{\textbf{Opacity in a single pulse: comparison with the semi-analytical work of \cite{granot:2008}.} 
$\gamma\gamma$ absorbed lightcurves at four different energies $\epsilon = E_\mathrm{HE}/ m_\mathrm{e}c^2$ are plotted as a function of the observer time for a single pulse using the prescriptions and model parameters corresponding to  the middle panel of Fig.~9  in \cite{granot:2008}. 
Our numerical calculation (solid line) is compared to the semi-analytical result of \cite{granot:2008} (dashed line). Notations are the same (the observer time $\overline{T}$ and observed fluxes $\epsilon F_\epsilon / F_0$  have normalized values). The agreement is excellent except for 
$\overline{T} < 10^{-2}$, where the discrepancy  is due to numerical resolution limitations (this corresponds to a true observer time $t_\mathrm{obs} < 0.1$ ms).  [figure from \cite{hascoet:2011}]}
\label{fig_granot}
\end{figure}

\noindent \textbf{General $\gamma$$\gamma$ opacity formula.} 
The $\gamma$$\gamma$ opacity ($\tau_{\gamma \gamma}$) is given by:

\begin{eqnarray}
 \tau_{\gamma \gamma} (E_{GeV}) & = & \int_{l_e}^{\infty} dl \int d \Omega \int_{E_c(E_{GeV}, \psi)}^{\infty} \nonumber \\
& & dE \ n_{\Omega}(E) \sigma_{\gamma \gamma}(E,\psi) (1-cos\psi)
\label{tau_gamma_general}
\end{eqnarray}

\noindent All the physical quantities are measured in the laboratory (or source) frame.
$E_{GeV}$ is the energy of the photon for which $\tau_{\gamma \gamma}$ is calculated whereas $E$ is the energy of the interacting field photon.
$\psi$ represents the interaction angle between the GeV photon and the interacting photon and $\sigma_{\gamma \gamma}$ is the $\gamma$$\gamma$ interaction cross-section between these two photons. 
$E_c = 2(m_e c^2)^2/[E_{GeV}(1-cos\psi)]$ is the energy threshold of the field photon above which $\gamma$$\gamma$ annihilation can happen.
Finally $n_{\Omega}$ is the photon field distribution $[\mathrm{ph} \cdot \mathrm{cm}^{-3} \cdot \mathrm{erg}^{-1} \cdot \mathrm{sr}^{-1}]$ at a given location and time.

\noindent The equation (\ref{tau_gamma_general}) is made of a triple integral : the $dl$-integration is done over the path of the GeV photon from its emission location to the observer, the $d \Omega$-integration is done over the solid angle distribution of the interacting photon field surrounding the GeV photon whereas the $d E$-integration is done over its energy distribution. The equation (\ref{tau_gamma_general}) is general and can be applied to any photon emitted at a given location and time with a given propagation direction within the GRB outflow.  \newline

\noindent \textbf{Validation of the model.} 
 The kernel of our study is the calculation of the $\gamma$$\gamma$ opacity created by a spherical flash, i.e. an instantaneous flash of photons emitted by an expanding relativistic spherical front. It is then possible to model the case of a propagating radiating spherical front (representing for example a shock wave) by the succession of many spherical flashes. One of the critical step is the exact calculation of the photon density $n_{\Omega}$ taking into account all the relativistic effects. Before dealing with more complex dynamical configurations within the internal shock framework, the validity of our numerical approach was tested on a simple single-pulse case with a comparison to the previous semi-analytic study of \cite{granot:2008} (see \reffig{fig_granot}).

%%%%%%%%%%%%%%%%%%%%%%%%%%%%%%%%%%%%%%%%%%
\vspace{-0.2cm}
\section{Application to Internal Shocks}
\subsection{Internal shocks within a relativistic outflow}
Now the model is applied to dynamical evolutions expected in the internal shock framework, where the whole prompt $\gamma$-ray emission is produced by electrons accelerated by shock waves propagating within a relativistic variable outflow.
We model the dynamics via a multiple shell model where the successive collisions between shells mimic the propagation of shock waves \cite{daigne:1998}. Each collision produces an elementary spherical flash: the simulated light curves are the result of the sum of all flashes. 
For each high energy photon, the $\gamma$$\gamma$ opacity is computed by integrating equation (1) from its emission location to the observer taking into account the exact radiation field $n_\Omega$ produced by all the collisions in the outflow.
A previous study of the $\gamma$$\gamma$ opacity in internal shock was made by \cite{aoi:2010}. However the prescription used to compute $\tau_{\gamma \gamma}$ was still approximate, using the local physical conditions of the outflow where the high energy photon is emitted and applying them to an average formula of $\tau_{\gamma \gamma}$ (as can be found in \cite{lithwick:2001, abdo:2009a, ackermann:2010}).

%%%%%%%%%%%%%%%%%%%%%%%%%%%%%%%%%%%%%%%%%%%
\subsection{Minimum Lorentz factor in GRB outflows -- The case of GRB 080916C}
\label{subsection_lfmin}
%%%%%%%%%%%%%%%%%%FIGURE%%%%%%%%%%%%%%%%%%%%%%%%%%%%%%%
\begin{figure*}
\begin{center}
\begin{tabular}{cc}
\includegraphics[scale=0.38]{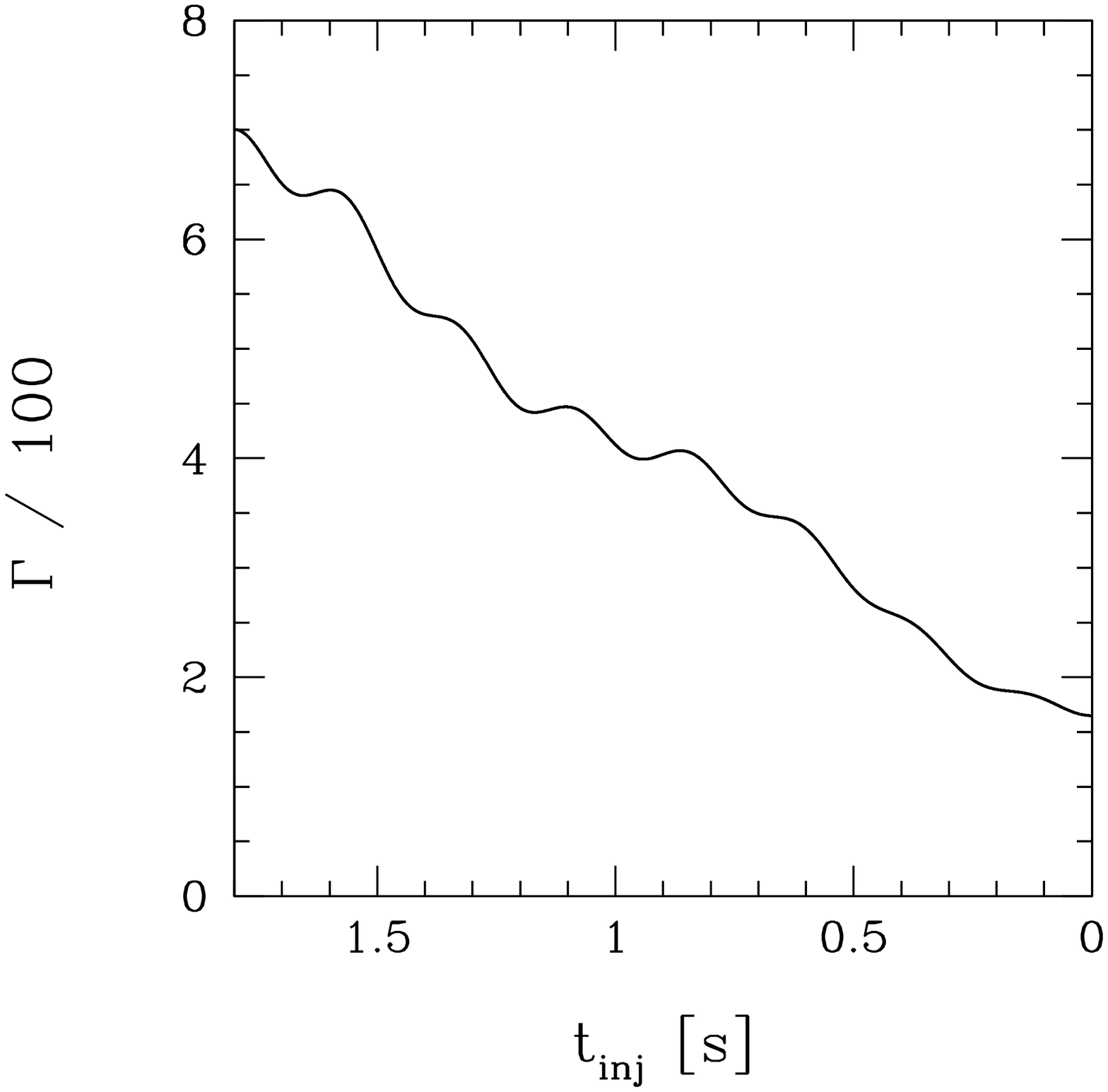} &  \includegraphics[scale=0.38]{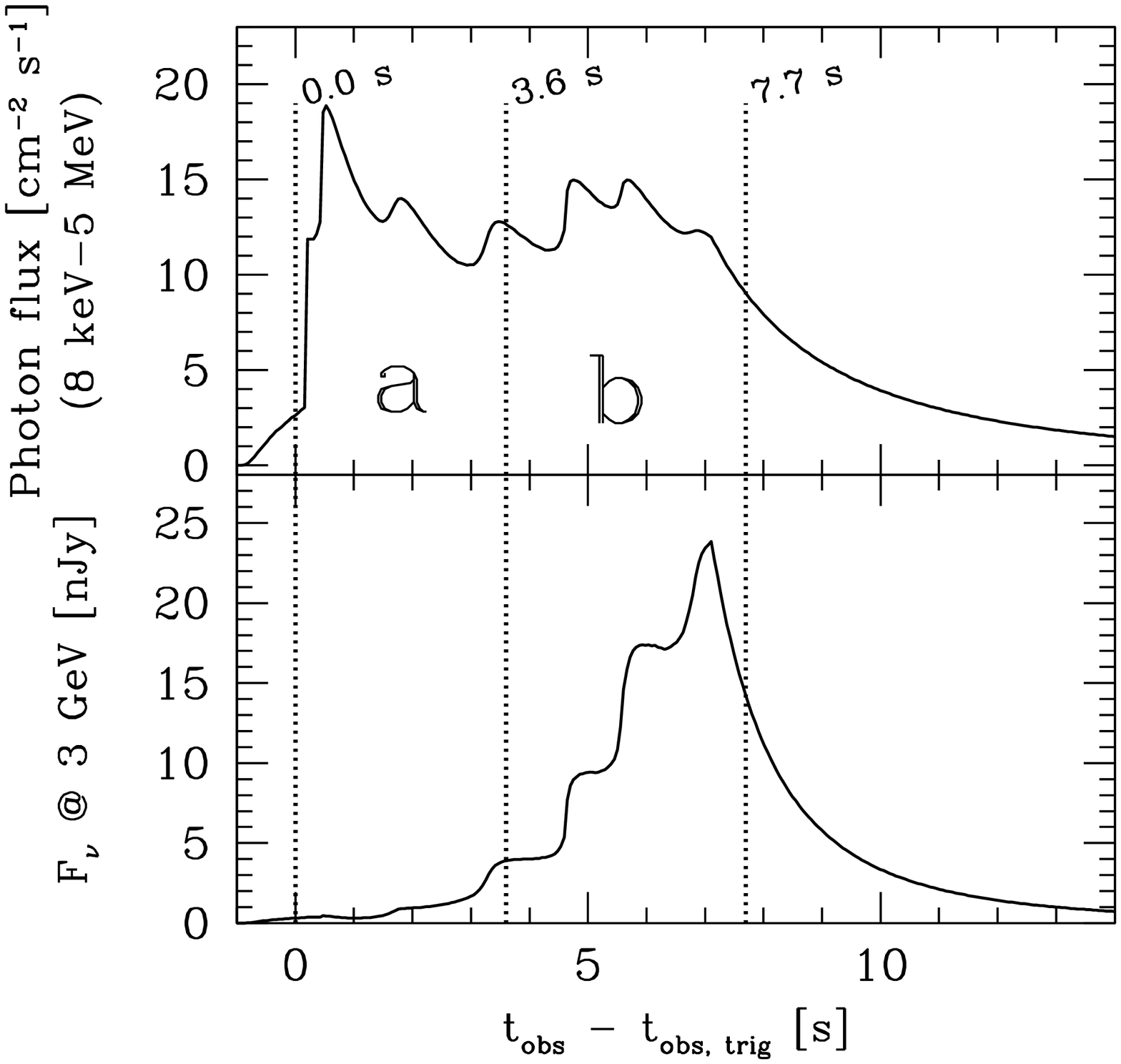} \\
\includegraphics[scale=0.38]{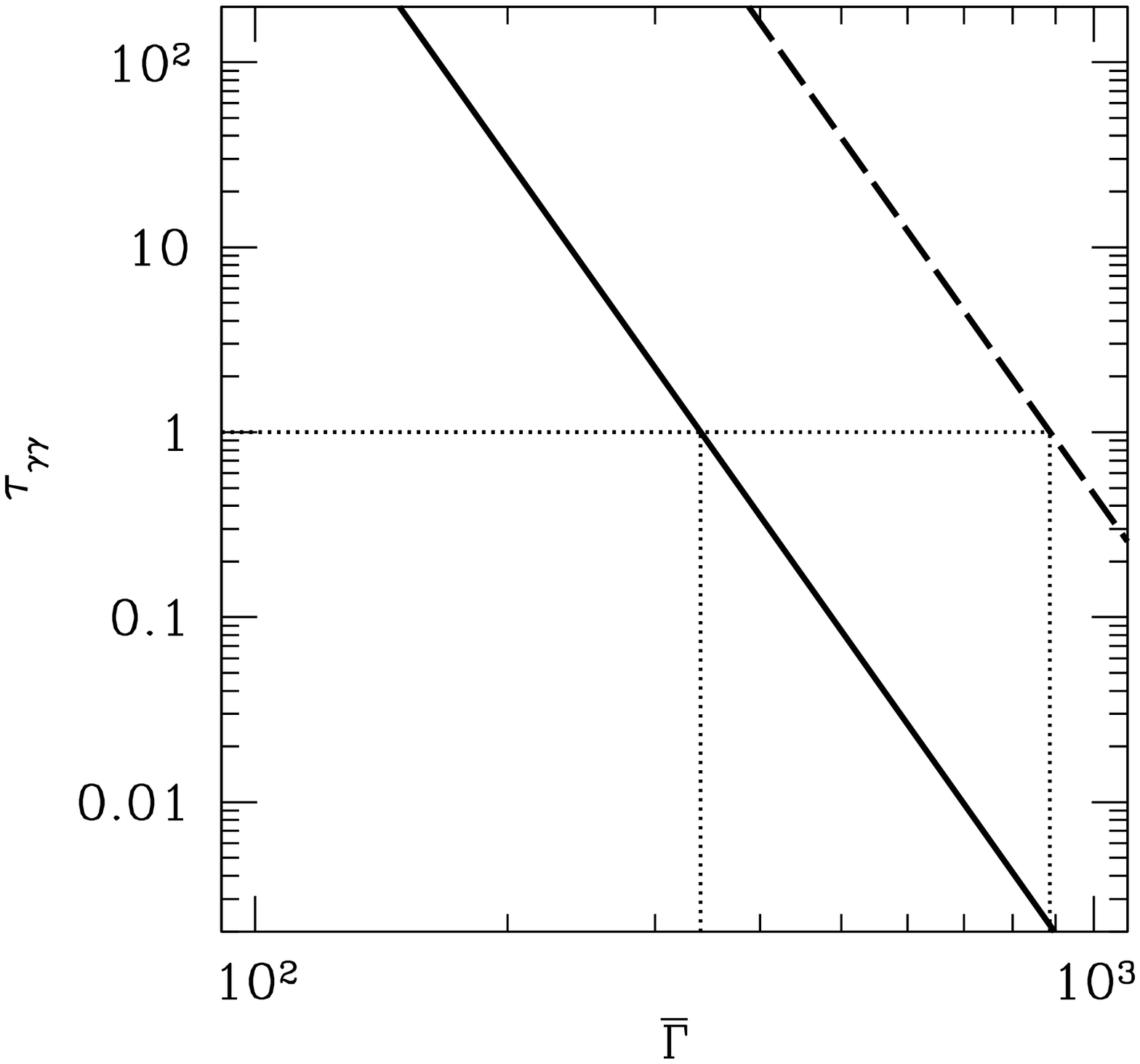} &  \includegraphics[scale=0.38]{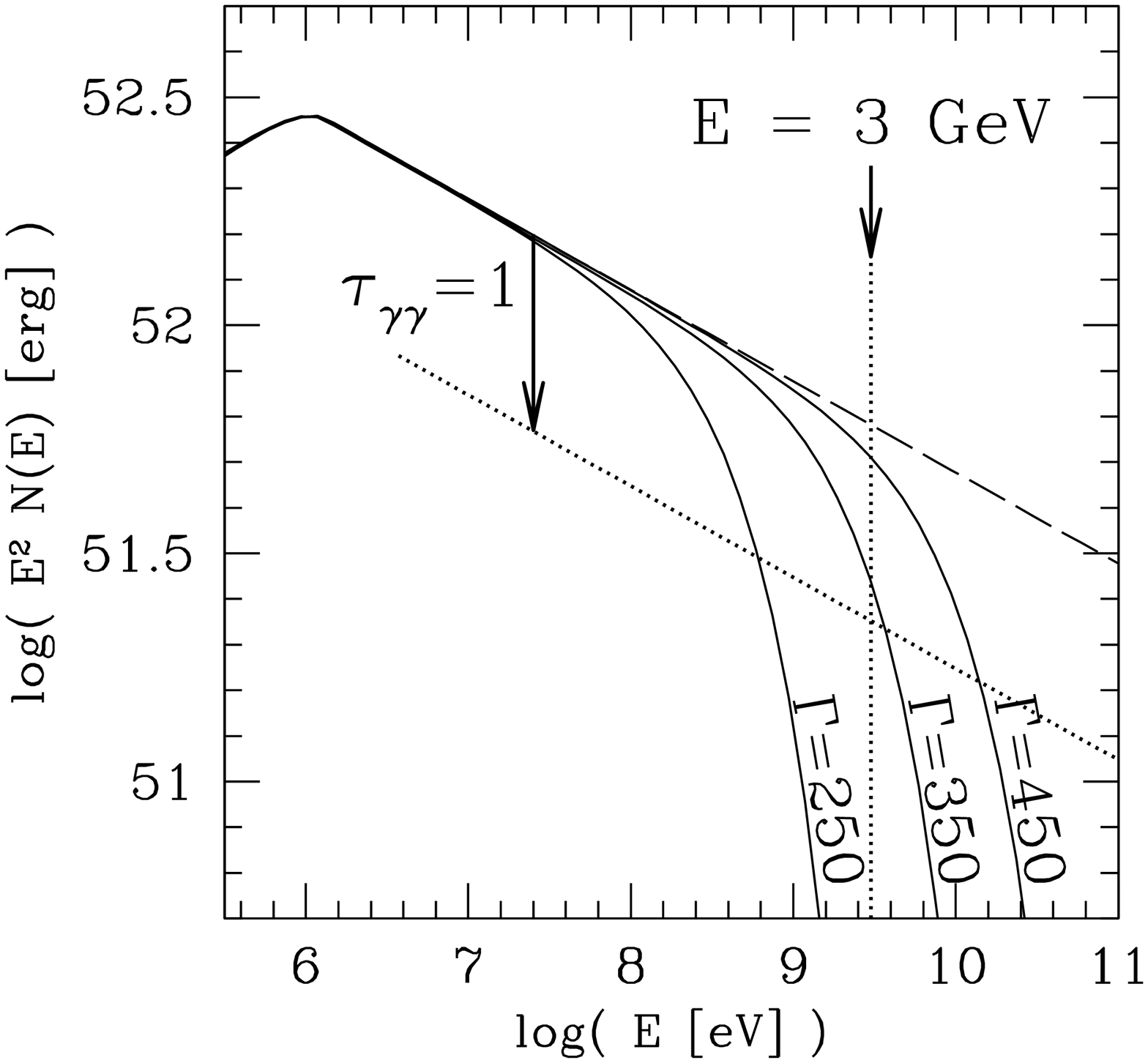}
\end{tabular}
\end{center}
\vspace*{-6ex}
\caption{\textbf{Minimum Lorentz factor for GRB 080916C.}  The two first panels are plotted for the limit case leading to $\tau_{\gamma\gamma}(3\,\mathrm{GeV})=1$ in time bin 'b', i.e. for a mean Lorentz factor $\overline{\Gamma} = \overline{\Gamma}_\mathrm{min}=340$. \textit{Upper left panel:} initial Lorentz factor distribution in the outflow. \textit{Upper right panel:} $\gamma$-ray lightcurves in the GBM band ($8$ keV -- $5$ MeV, top) and at 3 GeV (bottom). The lightcurves are plotted as a function of $t_\mathrm{obs}-t_\mathrm{obs,trig}$, where $t_\mathrm{obs,trig}$ is the observer time of the first detected photons. \textit{Lower left panel:} evolution of $\tau_{\gamma \gamma}$ at $E_\mathrm{HE}=3$ GeV against the mean Lorentz factor in the outflow $\overline{\Gamma}$, following our detailed modeling (solid line) and using the average formula from \cite{abdo:2009a} (dashed line). \textit{Lower right panel:} time integrated spectrum over time bin 'b' for different mean Lorentz factors  (the relative shape of the initial Lorentz factor distribution is kept the same) and reference spectrum without $\gamma\gamma$ annihilation (dashed line). [figure from \cite{hascoet:2011}] }
\label{fig_gmin}
\end{figure*}
%%%%%%%%%%%%%%%%%%FIGURE%%%%%%%%%%%%%%%%%%%%%%%%%%%%%%%
The first natural application of our model is the estimate of the minimum bulk Lorentz factor $\Gamma_\mathrm{min}$ in GRB outflows, obtained from the constraint $\tau_{\gamma\gamma}(E_\mathrm{HE,max})\simeq 1$, where $E_\mathrm{HE,max}$ is the highest photon energy detected in the burst. 
To illustrate this aspect with an example, we applied our approach to the case of one of the four brightest GRBs detected in the GeV range by \textit{Fermi}, i.e. GRB 080916C. The results are shown in \reffig{fig_gmin}. Using our numerical model, a synthetic GRB was generated, which reproduces the main observational features:
the total radiated isotropic $\gamma$-ray energy ($E_\mathrm{iso} = 8.8 \times 10^{54}$ ergs between 10 keV and 10 GeV), the spectral properties ($E_\mathrm{p}$, $\alpha$, $\beta$ parameters of the Band function
, the envelop of the light curve and a short time-scale variability of 0.5 s in the observer frame. The study is focused on the most constraining time bin (time bin 'b'),
during which the highest observed photon energy was $E_\mathrm{HE,max}=3$ GeV (16 GeV in the source rest frame): for this reason, only time bins 'a' and 'b' are reproduced in the synthetic GRB. These two intervals correspond to 32~\%
of the total radiated isotropic equivalent energy. The minimum mean Lorentz factor $\overline{\Gamma}_\mathrm{min}$ is obtained by requiring that  $\tau_{\gamma \gamma}\left(E_\mathrm{HE,max}\right) \le 1$ (see \reffig{fig_gmin}, lower panel). 
With the detailed calculation,  we find a minimum mean Lorentz factor $\overline{\Gamma}_\mathrm{min} = 340$, i.e. a factor $2.6$ lower than the value $\overline{\Gamma}_\mathrm{min}=887$, which was obtained from an approximate ``single zone'' model \citep{abdo:2009a}. Even more remarkable, the whole initial distribution of the Lorentz factor used in this model of GRB 080916C (from 170 to 700) remains below the ``minimum'' value of the Lorentz factor derived from single zone models (see  \reffig{fig_gmin}, upper left panel).

%%%%%%%%%%%%%%%%%%%%%%%%%%%%%%%%%%%%%%%%%%%
\subsection{Is the delayed onset of the GeV emission a signature of the $\gamma\gamma$ opacity ?}
\label{subsection_delay}
%%%%%%%%%%%%%%%%%%%%%%%%%%%%%%%%%%%%%%%%%%%
The high energy emission (above 100 MeV) detected by \textit{Fermi} in a few bright GRBs often shows a delayed onset compared to the softer $\gamma$-ray emission (below 5 MeV). 
The analysis by \cite{bbzhang:2011} indicates that such a delayed onset is present in at least 7 in a sample of 17 GRBs detected by \textit{Fermi}-LAT. This feature seems to be common to long and short GRB classes and its origin is debated \citep{granot:2010}. 
Among the proposed explanations (see e.g.\citep{zou:2009,li:2010,toma:2009}), the possibility that this delayed onset is induced by a $\gamma\gamma$ opacity temporal evolution effect has already been discussed by \cite{abdo:2009a}:  
as the shock wave producing the $\gamma$-ray emission expands to larger radii, the opacity seen by the  
high energy photons evolve from an optically thick to an optically thin regime. The model developed in the present study is well appropriate to investigate this possibility in more details. 
The synthetic burst used in \reffig{fig_gmin} to model bins 'a' and 'b' of GRB 080916C gives an example of a delayed onset at 3 GeV induced by an evolving $\gamma\gamma$ opacity. The first pulse is produced at lower radii and in lower Lorentz factor material and is therefore strongly absorbed. For this reason, it is almost suppressed in the 3 GeV lightcurve, whereas the second pulse is well visible. Note that the model reproduces simultaneously the onset delay of $\simeq 5 \ \mathrm{s}$ at high energy, and the short timescale variability of $\simeq 0.5 \ \mathrm{s}$ at low energy.

%%%%%%%%%%%%%%%%%%%%%%%%%%%%%%%%%%%%%%%%%%%
\vspace{-0.2cm}
\section{Consequences of distinct emission regions for MeV and GeV photons}
\label{sec_2zones_model}

\subsection{Are GeV and MeV photons produced in the same place ?}
It has been proposed in several recent studies that the delayed onset and/or the long-lasting tail of the high energy emission could be an evidence in favor of two different regions for the emission of MeV and GeV photons.
An extreme version is the scenario proposed by \cite{kumar:2010,ghisellini:2010}
where the whole GeV emission (prompt and long lasting) is produced by the external shock during the early deceleration of the relativistic outflow. Note that this scenario leads to strong constraints on the density and magnetization of the external medium \citep{piran:2010} and that the observed temporal slope of the long-lasting high-energy emission would imply a strongly pair-enriched medium \citep{ghisellini:2010}.

 Even in scenarios where the prompt GeV emission has an internal origin, a partially distinct emission region could be due to  a spectral evolution of the prompt mechanism. For instance, in the framework of internal shocks, the evolution of the physical conditions in the shocked region during the propagation of a shock wave leads to an evolving efficiency of the IC scatterings, depending on the importance of Klein-Nishina corrections. This naturally leads to a variable high-energy component following with a delay the main (Band) component in the MeV range \citep{wang:2009,bosnjak:2009,daigne:2011}. Successive generations of collisions in a variable outflow can also lead naturally to different emission regions \citep{li:2010}.
  An evolution in the microphysics of the acceleration process could also be responsible for some spectral evolution in scenarios where there is a dominant hadronic component at high energy (see e.g. \cite{asano:2009}). Finally, two emission regions are naturally expected in photospheric models, as it is often assumed that the main (Band) component has a photospheric origin and that internal shocks or magnetic dissipation occurring at larger distance produce an additional component at high energy (see e.g. \cite{toma:2010,vurm:2011}).

\subsection{Loosening the constraint on $\Gamma_\mathrm{min}$}
\label{sec_gmin_2zones}
As discussed in \cite{zhao:2011,zou:2011}, the possibility for the GeV photons to be produced in a different region than the MeV photons can loosen the constraint on the minimum Lorentz factor in GRB outflows. To investigate this effect,  
we consider the same synthetic GRB as used in \refpar{subsection_lfmin} to model time bins 'a' and 'b' of GRB 080916C. We focus on the onset of the GeV component, which occurs at $t_\mathrm{obs,onset}=t_\mathrm{obs,trig}+0.67\left(1+z\right)\,\mathrm{s}$, where $t_\mathrm{obs,trig}$ corresponds to the observer time of the first MeV photons. The MeV photons observed at $t_\mathrm{obs,onset}$ are emitted at radius $R_\mathrm{MeV}$ and it is assumed that the emerging GeV photons observed at the same time were emitted 
 by material moving with Lorentz factor $\Gamma_\mathrm{GeV}$ (velocity $\beta_\mathrm{GeV} c$)
at radius $R_\mathrm{GeV}$ and time $t_\mathrm{GeV}$ (source frame) with $t_\mathrm{GeV}-R_\mathrm{GeV}/c =  t_\mathrm{obs,onset}/(1+z)$. The flash of GeV photons emitted at $R_\mathrm{GeV}$ is assumed to have a power-law spectrum with photon slope $\beta=-2.2$. We define a latitude-averaged $\gamma\gamma$ opacity for GeV photons of energy $E_\mathrm{GeV}$ by
\begin{equation}
e^{-\overline{\tau}_{\gamma\gamma}\left(E_\mathrm{GeV}\right)} = \frac{ 
\int e^{- \tau_{\gamma\gamma}(E_\mathrm{GeV},\Theta_\mathrm{e})} \mathcal{D}(\Theta_\mathrm{e})^{1-\beta}  \sin{\Theta_\mathrm{e}} d \Theta_\mathrm{e}
}{ 
\int \mathcal{D}(\Theta_\mathrm{e})^{1-\beta}  \sin{\Theta_\mathrm{e}} d \Theta_\mathrm{e}
}\, ,
\label{eqn_mean_tgg}
\end{equation}
where $\tau_{\gamma\gamma}(E_\mathrm{GeV},\Theta_\mathrm{e})$ is the opacity seen 
photons emitted at colatitude $\Theta_\mathrm{e}$ and $\mathcal{D}\left(\Theta_\mathrm{e}\right)=\left( \Gamma_\mathrm{GeV}\left(1-\beta_\mathrm{GeV}\cos{\Theta_\mathrm{e}}\right)\right)^{-1}$ is the corresponding Doppler factor. The contribution of each colatitude to the mean value is weighted by the corresponding fluence, leading to the $1-\beta$ exponent.
%%%%%%%%%%%%%%%%%%%%%%%%%%%%%%%%%%%%%%FIGURE
\begin{figure}
\begin{center}
\begin{tabular}{c}
\includegraphics[scale=0.38]{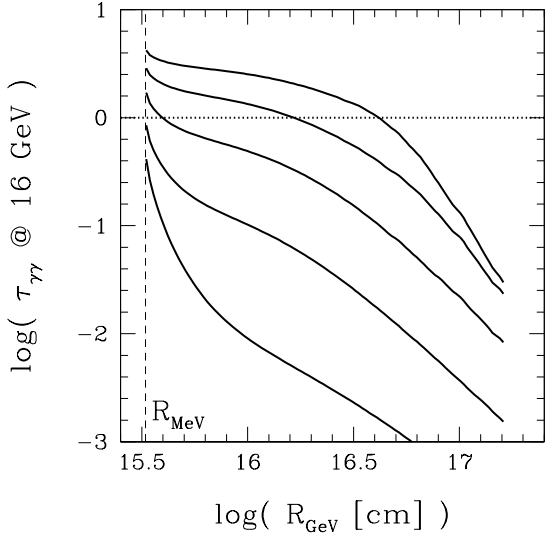} 
\end{tabular}
\end{center}
\caption{\textbf{Two emitting region scenario.} The $\gamma\gamma$ opacity $\tau_{\gamma \gamma}$ seen by 16 GeV  photons (source frame) observed at $t_\mathrm{obs,onset}-t_\mathrm{obs,trig}\simeq 0.67~(1+z)$ s (see text) is plotted as a function of their emission radius $R_\mathrm{GeV}$ for different values of the Lorentz factor  $\Gamma_\mathrm{GeV}=50$, $100$, $200$, $400$ and $800$ from top to bottom.  [figure from \cite{hascoet:2011}]}
\label{fig_2zones}
\vspace{-0.2cm}
\end{figure}
%%%%%%%%%%%%%%%%%%%%%%%%%%%%%%%%%%%%%%
We plot in \reffig{fig_2zones} the evolution of the latitude averaged $\gamma\gamma$ opacity $\overline{\tau}_{\gamma\gamma}$ at 16 GeV (source frame) as a function of $R_\mathrm{GeV}$ for $R_\mathrm{GeV}>R_\mathrm{MeV}$ and for different values of the Lorentz factor $\Gamma_\mathrm{GeV}$. When GeV and MeV photons are emitted at the same location, we find that $\overline{\tau}_{\gamma\gamma}\le1$ for $\Gamma_\mathrm{GeV} \ge \Gamma_\mathrm{GeV,min,same\,zone}\simeq 340$, i.e. the same limite as in \refpar{subsection_lfmin}. 
When $R_\mathrm{GeV}$ increases, the opacity $\overline{\tau}_{\gamma\gamma}$ decreases as expected, which loosen the constraint on the minimum Lorentz factor $\Gamma_\mathrm{min,GeV}$ of the material emitting GeV photons:
\begin{center}
\begin{tabular}{l||cccc}
$R_\mathrm{GeV}/R_\mathrm{MeV}$ & 1 & 1.2 & 5.1 & 13\\
\hline
$\Gamma_\mathrm{min,GeV}/\Gamma_\mathrm{min,GeV,same\,zone}$ & 1 & 0.59 & 0.29 & 0.15
\end{tabular}
\end{center}
This follows approximatively the dependency on $R_\mathrm{e}/R_0$ found in \cite{hascoet:2011} (section $2.2.3$), i.e. 
 $\tau_{\gamma\gamma}\propto \left(R_\mathrm{e}/R_0\right)^{2(\beta-1)}$ when $R_\mathrm{e}\gg R_0$, leading to $\Gamma_\mathrm{min,GeV}\propto \left(R_\mathrm{e}/R_0\right)^{-1}$.
As shown in \refpar{subsection_lfmin} the detailed modeling of the $\gamma\gamma$ opacity in a scenario where GeV and MeV photons are emitted in the same regions leads to a reduction of the minimum Lorentz factor by a factor $\simeq 2$--$3$ compared to single zone models. The calculation presented here shows in addition that the minimum Lorentz factor can be reduced further more by another factor $\simeq2-8$ for $\Gamma_\mathrm{GeV}$ if GeV emission becomes efficient at a radius larger than for MeV photons. Assuming that the radiated energy at $R_\mathrm{GeV}$ is not larger than the radiated energy at $R_\mathrm{MeV}$, we have checked that the outflow remains optically thin for the Thomson opacity due to primary electrons and secondary leptons at $R_\mathrm{GeV}$ in the case shown in \reffig{fig_2zones}.
Note that this result on the loosening of the constraint on the minimum Lorentz factor does not apply to models where GeV photons are entirely due to the external shock. Indeed, the small value of $t_\mathrm{obs,onset}$ implies an early deceleration. As the isotropic equivalent energy of GRB 080916C is huge, this leads to a minimum Lorentz factor $\overline{\Gamma} > 10^3$ in the outflow, which is more constraining that the $\gamma\gamma$ opacity limit.
The discussion of the effect of a distinct GeV emission region presented here is quite simplified and some limitations should be kept in mind. If an additional GeV component could be firmly identified in GRB 080916C, the maximum energy $E_\mathrm{MeV,max}$ of photons associated with the main component should be taken into account to derive a new constraint $\Gamma_\mathrm{min}$ on the Lorentz factor of the outflow during the MeV emission phase. We have assumed here $\overline{\Gamma}=340$ as derived in \refpar{subsection_lfmin} using $E_\mathrm{MeV,max}=16$ GeV (source frame) but $\Gamma_\mathrm{min}$ will be reduced
if $E_\mathrm{MeV,max}$ is lower. There is one further complication: the component produced at $R_\mathrm{GeV}$ extends probably in the soft gamma-ray range, as suggested by the observation of a soft excess correlated with the high energy component in some GRBs such as  GRB 090926 \citep{ackermann:2011}, GRB 090926 \citep{abdo:2009b} and GRB 090510 \citep{ackermann:2010}. It has been assumed here that the annihilation rate of GeV photons with the seed photons produced at $R_\mathrm{GeV}$ is negligible compared to the annihilation
rate with MeV photons produced earlier. This is however not necessarily the case depending on the relative intensity of the two components. 
Clearly, a detailed modeling of the emitted spectrum is necessary to investigate such effects. This is beyond the scope of these proceedings and we leave to a forthcoming study the coupling of the formalism presented here to compute the $\gamma\gamma$ opacity with a detailed radiative model such as developed by  \cite{bosnjak:2009}.

%%%%%%%%%%%%%%%%%%%%%%%%%%%%%%%%%%%%%%%%%%%
\vspace{-0.2cm}
\section{Conclusions}
\subsection{A new formula of $\Gamma_{min}$}
This study clearly illustrates the need for a detailed modeling to constrain the Lorentz factor in GRB outflows. However, when it is not possible, a reasonably accurate estimate of $\Gamma_\mathrm{min}$ can be obtained from the following formula
\begin{eqnarray}
\Gamma_\mathrm{min} & \simeq &  \frac{\left[C_1 2^{1+2\beta} \mathcal{I}(\beta)\right]^{\frac{1}{2(1-\beta)}}}{\left[\frac{1}{2}\left(1+\frac{R_\mathrm{GeV}}{R_\mathrm{MeV}}\right)\left(\frac{R_\mathrm{GeV}}{R_\mathrm{MeV}}\right)\right]^{1/2}}\, \left(1+z\right)^{-\frac{1+\beta}{1-\beta}}
\nonumber\\ 
& &\!\!\!\! \!\!\!\!\!\!\!\!\!\!\!\!\!\!\! \times
\left[ \sigma_\mathrm{T} \left(\frac{D_\mathrm{L}(z)}{c \Delta t_\mathrm{var}}\right)^2 \!\!\! E_\mathrm{c} F(E_\mathrm{c})\right]^{\frac{1}{2(1-\beta)}} \left(\frac{E_\mathrm{max}E_\mathrm{c}}{(m_\mathrm{e}c^2)^2}\right)^{\frac{\beta+1}{2(\beta-1)}}
\!\!\!\!
\, ,\nonumber\\ 
\label{eq_newapprox}
\end{eqnarray}
where $C_1 \simeq 4\cdot 10^{-2}$, $\Delta t_\mathrm{var}$ is the observed variability timescale,
$R_\mathrm{GeV}/R_\mathrm{MeV}$ is the ratio of the radii where the GeV and MeV components are emitted, 
and where the high energy spectrum (over a duration $\sim \Delta t_\mathrm{var}$) is assumed to follow a power-law with photon index $\beta$ above an observed characteristic energy $E_\mathrm{c}$ : $\overline{F}(E)=\overline{F}(E_\mathrm{c})(E/E_\mathrm{c})^\beta$ ($\mathrm{ph.cm^{-2}.keV^{-1}}$). Energy $E_\mathrm{max}$ is the observed energy of the most  energetic detected photons. 
As usually the spectrum is measured over a time interval $\Delta t_\mathrm{spec}$ which is larger than the variability timescale $\Delta t_\mathrm{var}$, the normalization $F(E_\mathrm{c})$ entering in \refeq{eq_newapprox} (fluence at energy $E_\mathrm{c}$ in $\mathrm{ph.cm^{-2}.keV^{-1}}$) must be corrected by a factor $F(E_\mathrm{c})=\overline{F}(E_\mathrm{c})\times \left(\Delta t_\mathrm{var}/\Delta t_\mathrm{spec}\right)$.
This equation can be directly applied to \textit{Fermi}-LAT observations and generalizes the usual formula given by \cite{abdo:2009a} by introducing two corrections: (1) a more accurate normalization including a numerical factor $C_1$ obtained from the comparison with numerical simulations presented in \refpar{subsection_lfmin} and \cite{hascoet:2011}; (2) the possibility to take into account two different emitting regions for MeV and GeV photons. The standard limit is obtained with $R_\mathrm{GeV}/R_\mathrm{MeV}=1$ (same region): then the denominator in \refeq{eq_newapprox} equals $1$. The radius $R_\mathrm{MeV}$ is estimated from the variability timescale by $R_\mathrm{MeV}\simeq \Gamma^2 c\Delta t_\mathrm{var}/(1+z)$, which is valid for most models of the prompt emission. The radius $R_\mathrm{GeV}$ is difficult to constrain without a detailed model of the high-energy emission mechanism. 
If GeV photons have an internal origin, an upper limit for $R_\mathrm{GeV}$ is given by the deceleration radius. In the future, a measurement of the variability timescale in the GeV lightcurve could provide a better estimate of this radius.

%%%%%%%%%%%%%%%%%%%%%%%%%%%%%%%%%%%%%%%%%%%
\vspace{-0.2cm}
\subsection{Other effects}
The detailed $\gamma$$\gamma$ opacity calculation model presented in these proceedings is appropriate and accurate to study many aspects and consequences of $\gamma$$\gamma$ annihilation in GRBs. In the present work we focus on the internal shock model and consider the consequences and signatures that $\gamma$$\gamma$ opacity could have in GRB observations, showing that: (i) the temporal evolution of $\tau_{\gamma \gamma}$ during a burst could favor a delay between the MeV and GeV light curves (ii) the $\gamma$$\gamma$ cutoff transition can be characterized in time-integrated spectra. It is usually closer to a power-law steepening than to a sharp exponential cutoff. The exact shape of the transition strongly depends on the details of the GRB dynamics. (iii) for complex GRBs, the $\gamma$$\gamma$ opacity could suppress the shortest time-scale features in high energy light curves (above 100 MeV). Only the point (i) is discussed in these proceedings: we refer the reader to the corresponding paper \cite{hascoet:2011} for more details on these different aspects.

%%%%%%%%%%%%%%%%%%%%%%%%%%%%%%%%%%%%%%%%%%%
\vspace{-0.2cm}
\begin{acknowledgments}
This work is partially supported by the French Space Agency (CNES).
R.H.'s PhD work is funded by a Fondation CFM-JP Aguilar grant.\newline
\end{acknowledgments}

%%%%%%%%%%%%%%%%%%%%%%%%%%%%%%%%%%%%%%%%%%%

\end{document}